\documentclass[conference]{IEEEtran}
\IEEEoverridecommandlockouts
\usepackage{cite}
\usepackage{amsmath,amssymb,amsfonts}
\usepackage{algorithmic}
\usepackage{graphicx}
\usepackage{textcomp}
\usepackage{xcolor}
\usepackage{url}
\usepackage{paralist}

\setdefaultleftmargin{1em}{1em}{1em}{}{}{}

\def\BibTeX{{\rm B\kern-.05em{\sc i\kern-.025em b}\kern-.08em
		T\kern-.1667em\lower.7ex\hbox{E}\kern-.125emX}}
\begin{document}
	
	\title{Decentralized Cross-Network Identity Management for Blockchain Interoperation}

	\author{\IEEEauthorblockN{Bishakh Chandra Ghosh\IEEEauthorrefmark{1}, Venkatraman Ramakrishna\IEEEauthorrefmark{2}, Chander Govindarajan\IEEEauthorrefmark{2}, Dushyant Behl\IEEEauthorrefmark{2},\\ Dileban Karunamoorthy\IEEEauthorrefmark{2}, Ermyas Abebe\IEEEauthorrefmark{2}, Sandip Chakraborty\IEEEauthorrefmark{1} }
		\IEEEauthorblockA{
            \IEEEauthorrefmark{1}\textit{Indian Institute of Technology Kharagpur}, \IEEEauthorrefmark{2}\textit{IBM Research}\\
			ghoshbishakh@iitkgp.ac.in, \{vramakr2, chandergovind, dushyantbehl\}@in.ibm.com,\\ \{dileban, ermyast\}@gmail.com, sandipc@cse.iitkgp.ac.in}
	
	}

		

	\IEEEoverridecommandlockouts
\IEEEpubid{\makebox[\columnwidth]{978-0-7381-1420-0/21/\$31.00~\copyright2021 IEEE \hfill} \hspace{\columnsep}\makebox[\columnwidth]{ }}
	
	\maketitle
	\IEEEpubidadjcol
	
	\begin{abstract}
		Interoperation for data sharing between permissioned blockchain networks relies on networks' abilities to independently authenticate requests and validate proofs accompanying the data; these typically contain digital signatures. This requires counterparty networks to know the identities and certification chains of each other's members, establishing a common trust basis rooted in identity. But permissioned networks are ad hoc consortia of existing organizations, whose network affiliations may not be well-known or well-established even though their individual identities are. In this paper, we describe an architecture and set of protocols for distributed identity management across permissioned blockchain networks to establish a trust basis for data sharing. Networks wishing to interoperate can associate with one or more distributed identity registries that maintain credentials on shared ledgers managed by groups of reputed identity providers. A network's participants possess self-sovereign decentralized identities (DIDs) on these registries and can obtain privacy-preserving verifiable membership credentials. During interoperation, networks can securely and dynamically discover each others' latest membership lists and members' credentials. We implement a solution based on Hyperledger Indy and Aries, and demonstrate its viability and usefulness by linking a trade finance network with a trade logistics network, both built on Hyperledger Fabric. We also analyze the extensibility, security, and trustworthiness of our system.
	\end{abstract}
	
	
	\section{Introduction}\label{sec:intro}

The current trend in enterprise blockchain industry is to create \textit{minimum viable ecosystems} for business consortia, i.e., limited business processes managed by permissioned blockchain networks that are built on diverse distributed ledger technologies (DLTs)~\cite{fabric, corda}. This has fragmented the wider blockchain ecosystem, producing isolated networks with data and assets in silos~\cite{abebe2019enabling}. But these networks have compelling reasons to interoperate while remaining operationally independent for privacy and logistical reasons. For example, an exported good that is financed on a trade finance network~\cite{wetrade, marcopolo} may be tracked on a separate provenance network~\cite{ibmfoodtrust} or a trade logistics network~\cite{tradelens}, while the traders' identities are attested on a shared KYC (Know Your Customer) network~\cite{kyc}. Such cross-network operations should be as trustworthy as transactions within those networks.
The problem of transferring or exchanging assets across networks is well-studied, especially in the context of permissionless networks~\cite{htlc, tesseract, herlihy2018atomic, xclaim}. We focus instead on the data sharing problem, that is, the transfer of data recorded in the ledger of one permissioned network to the ledger of another. To prevent fraud, the information must be accompanied by proof of veracity and provenance, as there is no guarantee that parties interested in it will be privy to both the ledgers. 

\begin{figure}
\centering
 \includegraphics[width=0.35\textwidth]{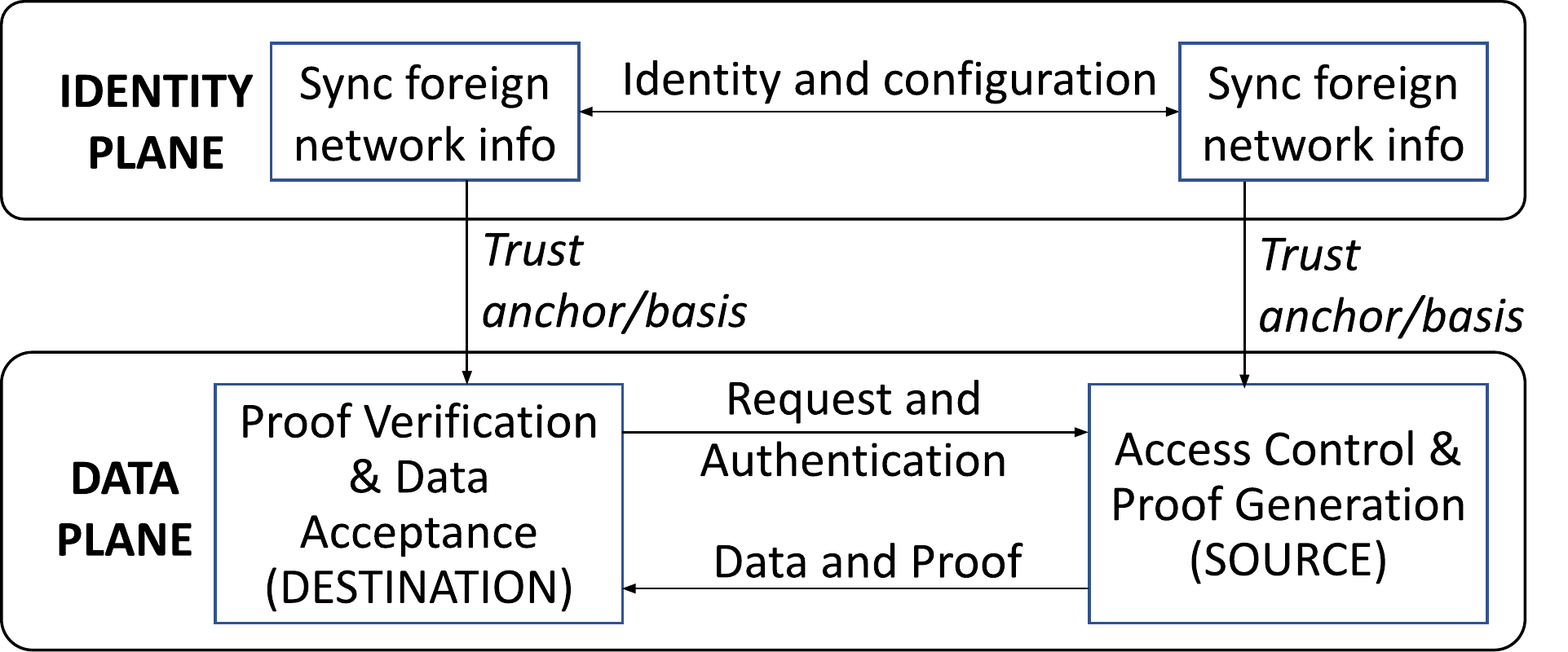}
 \caption{Generalized Cross-Network Data Transfer Protocol}
 \label{fig:high-level-protocol}
\end{figure}

Though several schemes exist for data sharing with proof, they are typically relevant for permissionless networks~\cite{popow, pegs}, rely on intermediary infrastructure (relay chains)~\cite{cosmos, polkadot}, or use API-level integration between trusted end-points. Because we want networks to remain autonomous and interoperate directly as per need, we base our work on the foundation laid by Abebe \textit{et al}~\cite{abebe2019enabling}. According to their model, as illustrated in the lower half (or \textit{data plane}) of Figure~\ref{fig:high-level-protocol}, a source network receiving an information request applies an access control policy through consensus before generating data and proof, and a destination network accepts the information after validating the proof against a verification policy, again through consensus. 
This proof must reflect the \textit{consensus view} of the source network's peers. Though the protocol is agnostic of the nature of proof and policy, the opacity of a permissioned network necessitates some form of \textit{proof-by-attestation}; i.e., the proof must consist of a quorum of network participants' signatures attesting to the veracity of shared data.
But proofs-by-attestation rely on a network's ability to gain some visibility into its counterparty network, to enable its participants to know the identities and certificate chains of the latter's participants so signatures in proofs can be validated. Gaining such visibility is therefore crucial to establish a trust basis for cross-network data sharing. In this paper, we separate the data requests and proof validation concerns from identity concerns by placing them in separate planes of activity (see Figure~\ref{fig:high-level-protocol}). Further, we replace the assumption made by Abebe \textit{et al} that the networks' participants have a priori knowledge of each others' identities and certificates, which is impractical to guarantee, with a generic and pluggable \textit{identity plane} protocol that provides a trust basis for data plane interoperation. In a naive implementation, credentials could be directly exchanged via network proxies, but this is both insecure (reliance on intermediaries) and unsustainable (discovering and exchanging credentials on the fly). Our primary contribution in this paper is the design of a secure distributed identity management infrastructure and set of protocols linking permissioned networks and laying the basis for blockchain interoperation.

To build such infrastructure, we rely on the fact that enterprise blockchain networks are created by mutual agreement among existing real-world organizations possessing digital identities, though their network affiliations are not well-attested outside the network consortia. We handle heterogeneity challenges in sharing and proving identity (credential formats, digital signature algorithms, sharing policies, identity providers) as well as privacy constraints (an organization may wish to reveal its network affiliation for cross-network data sharing while keeping its other attributes and affiliations secret). Our infrastructure enables networks to dynamically discover and sync each others' membership lists and verify the pre-existing identities of participant organizations without mandating a single centralized identity and validation registry (which would undermine the principle of decentralization and constrain the autonomy of interoperating networks). Dynamic changes in network memberships are handled, avoiding situations where non-participants or ex-participants can claim to be present participants of a network (which would result in invalid proofs of data). Finally, our system is agnostic of the DLT on which a network is built (structure, consensus protocol, identity issuance, cryptographic mechanisms).

In Section~\ref{sec:decentralized_group_identity_management}, we elaborate on the challenges and requirements while making the case for building our solution on the open frameworks like decentralized identifiers, verifiable credentials, and DLT-based identity registries. Our solution, built on Hyperledger Indy~\cite{indy} and Aries~\cite{aries}, with its building blocks, architecture, and protocols, are described in Section~\ref{sec:solution}, and a proof-of-concept implementation for Hyperledger Fabric~\cite{fabric} networks is presented in Section~\ref{sec:implementation}. We analyze our system in Section~\ref{sec:analysis} and survey related work in Section~\ref{sec:relatedork} before making concluding remarks in Section~\ref{sec:conclusion}.
	\section{Decentralized Group Identity Management}\label{sec:decentralized_group_identity_management}

An identity plane protocol involves distributed management of group identities. This is because a permissioned blockchain network is a collective rather than a unitary entity, deriving its identity from its participants (typically organizations), who may join or leave the network at any time. For privacy and security, such networks are open to outsiders only through invitation, and they manage identities internally in diverse ways, making cross-network identity management a challenge.
For example, in a Hyperledger Fabric network, each participating organization runs one or more \textit{Membership Service Providers (MSPs)} to issue identities and certificates to its peers and clients~\cite{fabric, fabric-msp}. Whereas in a Corda~\cite{corda} network, identity is managed through a hierarchy of certificate authorities (CAs), from a single root CA to one or more doormen CAs down to individual node CAs.

Bridging the identity gap between networks so that they can share data and validate proofs, therefore requires cross-network identity management. This can be done in different ways, but if we wish to let the networks remain autonomous, avoid dependence on centralized identity providers, and preserve blockchain tenets of consensus and distributed trust, we must overcome several technical challenges:

\begin{compactitem}
    \item \textbf{Platform heterogeneity:} networks built on different DLTs have different structures and different procedures for transaction commitment and consensus.
    \item \textbf{Identity management heterogeneity:} DLTs have diverse mechanisms for identity management and use diverse cryptographic algorithms.
    \item \textbf{Lack of common identity infrastructure:} external identity providers for network's participants (members), who may serve as a common root of trust, are themselves diverse (i.e, non-standardized), and there is also no guarantee that two networks will have a common set of identity providers to vouch for the members of both.
    \item \textbf{Privacy:} participants must be required to share only the bare minimum information necessary for interoperation.
    \item \textbf{Security:} outsiders and ex-members should not be able to claim that they belong to a given network; i.e., the integrity of network membership should be guaranteed.
    \item \textbf{Consensus on identity:} registering the identity of a foreign network's member organization in one's ledger creates a shared truth for a network and therefore must be governed through its consensus mechanism.
\end{compactitem}

To address these challenges, and knowing both the nature of permissioned networks and what technologies exist for distributed identity management, we can derive certain design requirements for our solution:
\\
\textbf{Decoupling Participants' Identities from Network:} Identities created within permissioned blockchain networks, e.g. Fabric root certificates, are unknown outside network boundaries. Since interoperation requires identities to be shared and validated externally, and to avoid creating a central authority or spokesperson, identities ought to be independently verifiable by external entities while remaining under the participants' control.
This necessitates the \textit{decoupling} of participants' identities from the networks they belong to. Fortunately, we can rely on the \textit{self-sovereign identity (SSI)} concept~\cite{ssi_inevitable, did}, which allows an entity to control/manage its identity and prove properties about itself while retaining independence from any centralized registry/provider.
\\
\textbf{Decentralized Identity Registries:} To complement SSI (as a means of decoupling), there must exist external registries to facilitate the resolution of network participants' decentralized identities~\cite{did} and the issuance, validation, update, and deactivation of credentials. To aid in mapping SSI to network-specific identity, such registries must maintain publicly accessible credential structures and validation information on behalf of one or more trusted identity-providing or credentialing authorities.
Several \textit{Decentralized Identifier} (DID) registries exist~\cite{didregisteries}, but our use cases ideally require those that are not governed by centralized authorities.
\\
\textbf{Maintaining Network Membership Integrity:} At any given instant, a network's membership list must be unambiguously available to a counterparty network in an interoperation session, and any changes to membership (joins and leaves) must be communicated 
using a trustworthy process. 
Incorrect membership information, or worse, malicious outsiders and ex-members pretending to belong to a network, will corrupt a data-sharing procedure that relies on proof-by-attestation, because a proof consumer may mistakenly accept an out-of-date (or fake) set of identities and signatures. 
Our solution must ensure that the membership information can be
validated though consensus in the receiving network at any given instant. 
\\	
\textbf{Trust Anchors to Discover and Certify Network Consortia:} Though a permissioned network is a voluntary consortium of independent members, once created, it acquires a life of its own and remains wedded to its identity even if most of the original participants leave. Hence, for a newly-formed network, it is more straightforward to discover its identity and subsequently discover its membership rather than deal with the chicken-and-egg problem of extrapolating a network identity from a group of participants' identities. For discovery, we must rely on reputed \textit{trust anchors} that may either directly represent the consortium or serve as an authoritative reference for it. Examples: Being well-known stakeholders, IBM or Walmart could represent (anchor) the IBM Food Trust Network~\cite{ibmfoodtrust}, whereas Maersk could represent Tradelens~\cite{tradelens}. Such anchors can be implemented with different levels of decentralization; they could be single entities or sets of corroborating entities representing different network participants; they could run their own organizations or belong to shared identity registries, maintaining credentials in shared ledgers. Ultimately, the trust anchor(s) should represent the collective rather than a single network participant to ensure decentralization.
\\	
\textbf{Compatibility with Networks' Identity Management Systems:} No identity plane protocol we develop ought to require any internal modifications to underlying blockchain platforms. Though these networks may have to implement additional adapters to manage heterogeneity for interoperation purposes, the internal transaction commitment and consensus processes, which rely on existing certification and cryptographic mechanisms, must be allowed to proceed unchanged. Therefore, our solution
must be simultaneously compatible with existing permissioned DLT identity management systems and agnostic of their specific implementations.

	\section{Solution}\label{sec:solution}

We now present a decentralized identity management solution to fulfill the requirements stated in Section~\ref{sec:decentralized_group_identity_management}, and whose development was guided by the following design principles:

\begin{compactitem}
\item The solution should not be tied to, or only applicable for, a particular DLT.
\item Networks and their participants should be free to choose identity registries and providers (or use their existing ones).
\item Networks must retain their autonomy while gaining the ability to interoperate universally.
\item No change should be required in a network's regular operation, nor should it be burdened with onerous additional configurations.
\end{compactitem}

\begin{figure}
	\centering 
	\includegraphics[width=0.9\linewidth]{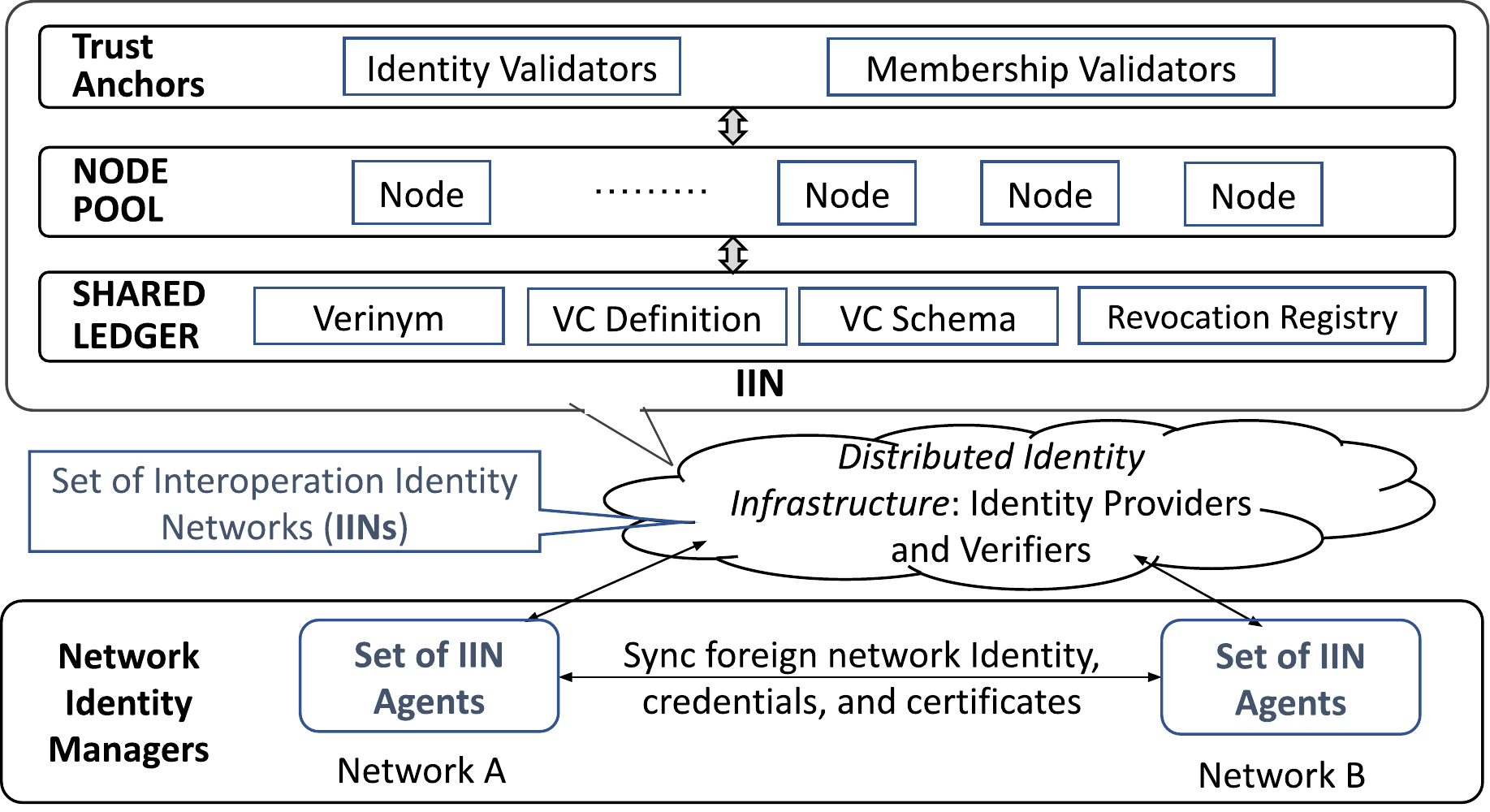}
	\caption{Architecture to enable identity plane exchanges}
	\label{fig:architecture}
\end{figure}

\subsection{Building Blocks}
We rely on existing decentralized identity management concepts and tools to serve as building blocks for our solution.

\begin{compactitem}
\item \textbf{Decentralized Identifiers (DIDs)} is a W3C draft~\cite{did} for self-sovereign identity (SSI). A DID is a URI that resolves to a DID Document that contains information about its subject like aliases and pseudonymns. It can contain public keys to authenticate the subject's signature and service endpoints for communication with the subject to obtain verifiable credentials (see further below). We assume each network participant possesses a DID, decoupling its external identity from its network affiliation. 

\item \textbf{Verifiable Credentials (VC)} is W3C specification on cryptographic digital credentials based on DIDs, used to certify claims about their holders\cite{vc}, who can then prove their claims to third parties using \textbf{Verifiable Presentations (VP)}.

\item \textbf{Distributed Verifiable Data Registry (VDR)} is a data registry that maintains identity records in a distributed ledger. Examples are Hyperledger Indy~\cite{indy} and Sidetree~\cite{sidetree}. Indy, built on a blockchain network, maintains DID records through consensus. (\textit{Note}: our design will accommodate centralized DID registries too but we recommend Indy-like networks for optimal levels of decentralization.)

\item \textbf{Identity and Credential Messaging} in a platform-neutral and interoperable protocol is required for peer-to-peer exchange of DID records, VCs, and VPs among issuers, subjects and verifiers. An example (though our design can accommodate any equivalent) is the \textit{DIDComm protocol} in \textit{Hyperledger Aries}~\cite{aries}, which facilitates such exchanges with support for encryption using keys and service endpoints stored in registry DID records.

\end{compactitem}

\begin{figure*}[t]
\centering
 \includegraphics[width=\textwidth]{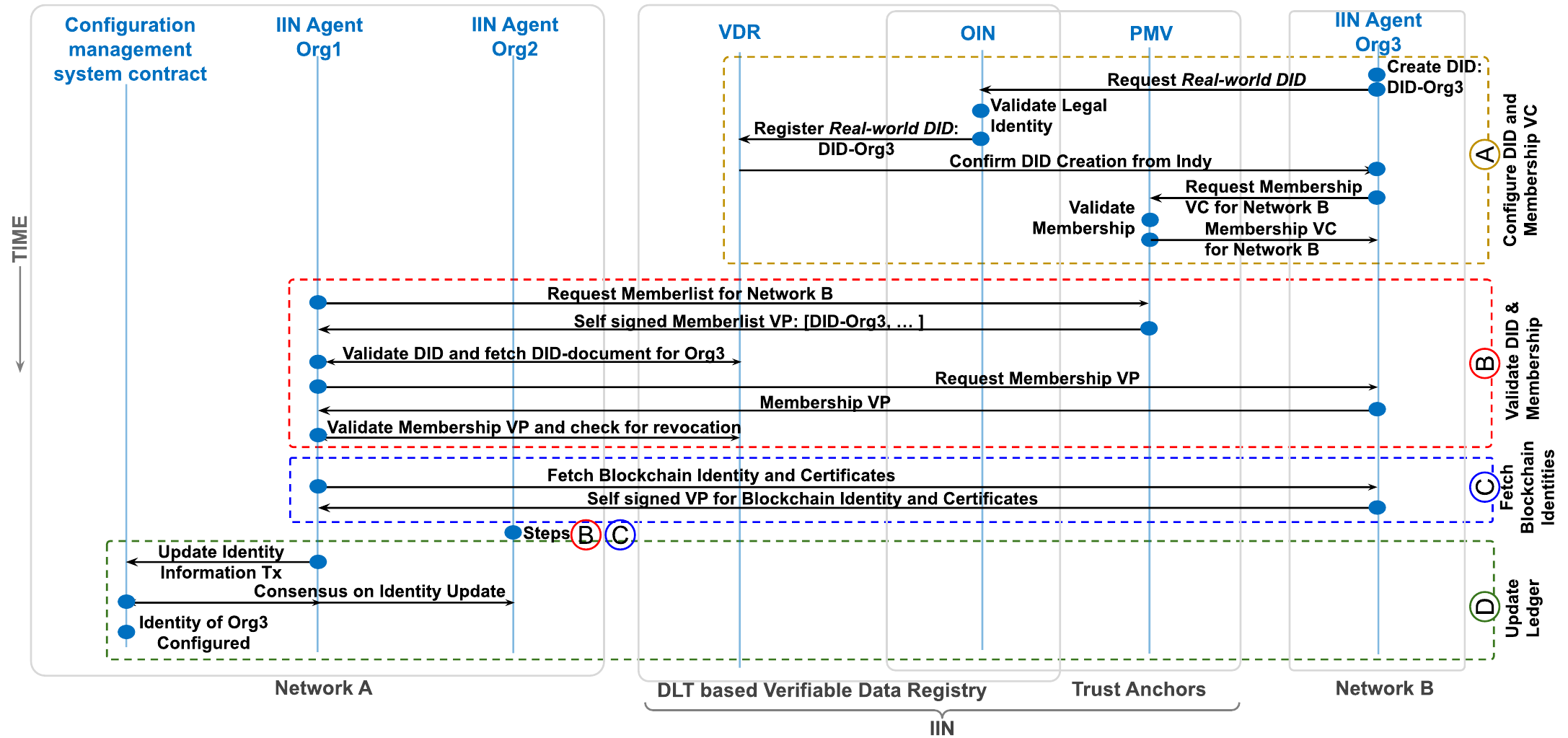}
 \caption{Identity Exchange Protocol}
 \label{fig:protocol}
\end{figure*}

\subsection{Architecture}

Two sets of modules are required to realize an identity plane protocol (Figure~\ref{fig:architecture}) - (i) a set of networks separate from the interoperating networks, collectively called \textit{Distributed Identity Infrastructure}, in which identity and credential records are maintained, and (ii) a set of agents within an interoperating network, collectively called \textit{Network Identity Managers}, each acting on behalf of a participant, syncing and validating identities across network boundaries, and recording foreign identities on the local ledger. 

\subsubsection{\textbf{Distributed Identity Infrastructure}}
This is a \textit{cloud} of what we term \textbf{Interoperation Identity Networks (IINs)}, which collectively provide a common root of trust for networks to sync identities and certificates. Each IIN consists of a distributed VDR, which, without loss of generality, is built on a public permissioned ledger allowing open queries but restricting writes to designated \textit{trust anchors} (\textit{see below}). The VDR ledger, shared and maintained by the IIN's pool of nodes through consensus~\cite{indy, aublin2013rbft}, also records verifiable credentials' schemas, authentication keys for credentials' attributes, and revocation lists. Each IIN in our infrastructure cloud enables trustworthy validation of certain networks' memberships by maintaining DID records for their participants and schemas and keys for issuing and verifying credentials.

\textbf{Trust Anchors}: An \textit{IIN} does not have a centralized source of trust. Instead, a trust basis is collectively created by \textit{trust anchors (TAs)}, entities that possess ledger write privileges, allowing them to issue DIDs to entities and maintain DID records on the IIN ledger, and further, issue VCs to DID owners. Some TAs are bootstrapped into an IIN, but any other DID owner can be assigned anchor privileges by an existing anchor too, thereby creating a trust hierarchy. Each TA in an IIN vouches for certain network members' real-world identities and attests to their participation (membership) in interoperating networks. A TA, in effect, is like a conventional identity provider but in our architecture, maintains identity records in a shared ledger with other TAs.

IIN TAs come in two varieties.
\textit{\textbf{Organization identity validators (OINs)}}, who already possess well-known real-world identities, are responsible for associating a network participant's DID with its real-world identity on an IIN ledger (through consensus). (\textit{Note}: The W3C draft proposal does not mandate a DID's automatic association with a real-world or legal identity upon creation~\cite{did,didusecase}, so we need OINs to make such associations explicit.) The presence of a DID record for a network participant with such an association implies that its identity is vouched for by one or more TAs of that IIN.
\textit{\textbf{Participant membership validators (PMVs)}} are responsible for validating the membership of a DID owner in a given permissioned network. Enterprise blockchain networks formed by mutual agreement among its participants have no single central authority that can certify the network's structure or its members. Therefore, proving participation of an organization requires attestation by one or more PMVs, which, like OINs, are reputed in industry or well-known as consortia representatives. For example, either IBM or Walmart, both reputed entities, could act as validators for the membership of the IBM Food Trust \cite{ibmfoodtrust} network, in whose launch and governance they both played key parts.
These trust anchors issue verifiable credentials to participants attesting their network memberships and also revoke them when organizations leave networks.
An organization may hold multiple VCs attesting its memberships in multiple networks. In a cross-network data sharing session, it can construct a VP proving its membership in either counterparty network without revealing its memberships in other networks, thus ensuring privacy \cite{vc}.

\textbf{IIN Artifacts:} the following are maintained on the legder:

\begin{compactitem}
\item \textit{Real-world DIDs} attesting the real-world identity of an organization (network participant).

\item \textit{Membership VC schema and authentication key} - A membership VC for an organization proves its participation in a given network. The VC contains its DID and the name/identity of a network it is a member of as attributes (claims). This \textit{Membership VC schema} is recorded on the IIN ledger, and every organization's VC must adhere to it, though different VCs may use different encodings and cryptographic algorithms. The public key used for authentication of this VC is also recorded on the ledger and used to validate membership claims made by the organization using a VP.

\item \textit{Memberlist VC schema and authentication key} - A memberlist VC consists of a name/identifier for a given network and a list of the network's participants' DIDs as schema attributes. Memberlist VCs are issued by PMVs and used by interoperating networks to fetch each others' list of participants, after which each participant's membership VC can be fetched and validated.

\item \textit{Revocation Registries} - When a blockchain network's member leaves, the \textit{Membership VC} indicating its affiliation with the network must be revoked. We use cryptographic accumulators~\cite{accumulators} as revocation registries, allowing membership presence checks without revealing the entire list of members. Each PMV registers a separate revocation registry on the IIN ledger. This registry is updated when a VC is revoked, and is also looked up by entities validating a verifiable presentation made by a Membership VC holder.

\end{compactitem}

\subsubsection{\textbf{Network Identity Managers}}
These components lie within the trust boundary of a permissioned network, one or more acting on behalf of each participating organization. A network identity manager is responsible for identity-syncing, i.e., (i) presenting its own identity and membership credentials to a foreign network, and (ii) correspondingly validating the membership credentials of a foreign network's members, and fetching and storing their certificates in the local ledger for data plane interoperation. We will henceforth refer to these managers as IIN Agents as they rely on IINs and their trust anchors for discovery and connections with foreign networks.

\textbf{IIN Agent:} This is responsible for registering the real-world DID of a network participant on an IIN and obtaining a Membership VCs from a TA of that IIN. It communicates with IIN Agents of foreign network participants using a confidential web-based channel to prove its own identity and validate their identity and membership claims using VPs.
Furthermore, \textit{IIN Agents} exchange their network-issued identity and certificates using self-signed VPs that can be verified against their real-world DID. Once verified, they configure these certificates in their network's shared ledger through consensus.
This maps an organization's decoupled SSI (real-world DID) to its network-issued identity, which ultimately makes proof verification possible in the data plane for interoperation.

\textbf{Ledger Artifacts:} Each permissioned network maintains the following policy configuration for identity-sharing, trust, and interoperation:

\begin{compactitem}
\item \textit{Interoperation network list:} list of foreign networks with which the local network is willing to interoperate.
\item \textit{Trust list:} IINs and specific TAs within that are trusted for identity and membership validation of foreign networks.
\item \textit{Foreign network identities:} identities of organizations participating in foreign networks and their network-issued credentials (typically certificate chains).
\end{compactitem}

\subsection{Identity Exchange Protocol}
\label{subsec:messageflow}
Figure \ref{fig:protocol} illustrates the steps in our canonical identity plane protocol to discover and sync identity and credentials across an example pair of networks (Network A and Network B) and a single IIN. Here, Org1 and Org2 in Network A are learning about the identity and membership of Org3 in Network B so the networks can share ledger data with each other. Note that by repeating these steps, identity information of any of the other organizations in Network B can be discovered, fetched, validated and configured in Network A.

\textbf{\textit{(A) Configure DID and Membership VC:}} Org3 creates a DID and requests an OIN to attest its identity and register it as a real-world DID.
Once this DID is registered in the IIN, Org3 must get a Membership VC issued to it by a PMV of that IIN. The PMV validates the request using some out-of-band validation procedure, issues the VC, and updates the revocation registry accordingly.
\textbf{\textit{(B) Validate DID \& Membership:}} Before starting the validation process, Org1 and Org2 need to know who the participants of Network B are. Org1 requests a Memberlist VC from the PMV associated with Network B, which returns a self-signed VP. After Network B's participants' real-world DIDs are known, Org1 resolves Org3's DID Document from the IIN ledger. This DID Document contains the service endpoint which is then used by Org1 to request Membership VP from Org3. Org1 then validates the VP received using the Membership VC schema, authentication key, and the revocation list, all fetched from the IIN ledger.
\textbf{\textit{(C) Fetch Blockchain Identity Information:}} Org1 requests Org3 for its network-issued identity and certificates, which Org3 returns in the form of a self-signed VP that is validated against Org3's real-world DID's authentication key.
\textbf{\textit{(D) Update Identity in Ledger:}} Though Org1 now has verified the identities of Org3, it cannot record it on Network A's ledger without a consensus among the network's participants. Therefore, Org2 independently carries out steps (B) and (C) above and endorses Org1's request to commit Org3's identity to the blockchain (using a smart contract transaction). Thus, no single participant of a network can unilaterally manipulate the local record of the identity of a foreign network's participant.

	\section{Use Case for Hyperledger Fabric}\label{sec:implementation}

We demonstrate a proof-of-concept implementation of our protocol by augmenting the two-network use case in Abebe et al \cite{abebe2019enabling}. We started with the already developed scaled-down versions of the trade logistics network, TradeLens~\cite{tradelens}, and the trade finance network, We.Trade~\cite{wetrade}, namely \textit{Simplified TradeLens (STL)} and \textit{Simplified We.Trade (SWT)} respectively. Here, each network  
runs Hyperledger Fabric peers, membership service providers (MSP), an ordering service, and application (client-layer) components, in Docker containers. Each network was equipped with a relay and two system contracts: Configuration Management and Data Acceptance Chaincode (CMDAC) and Exposure Control Chaincode (ECC).

\begin{figure}[t]
 \includegraphics[width=0.48\textwidth]{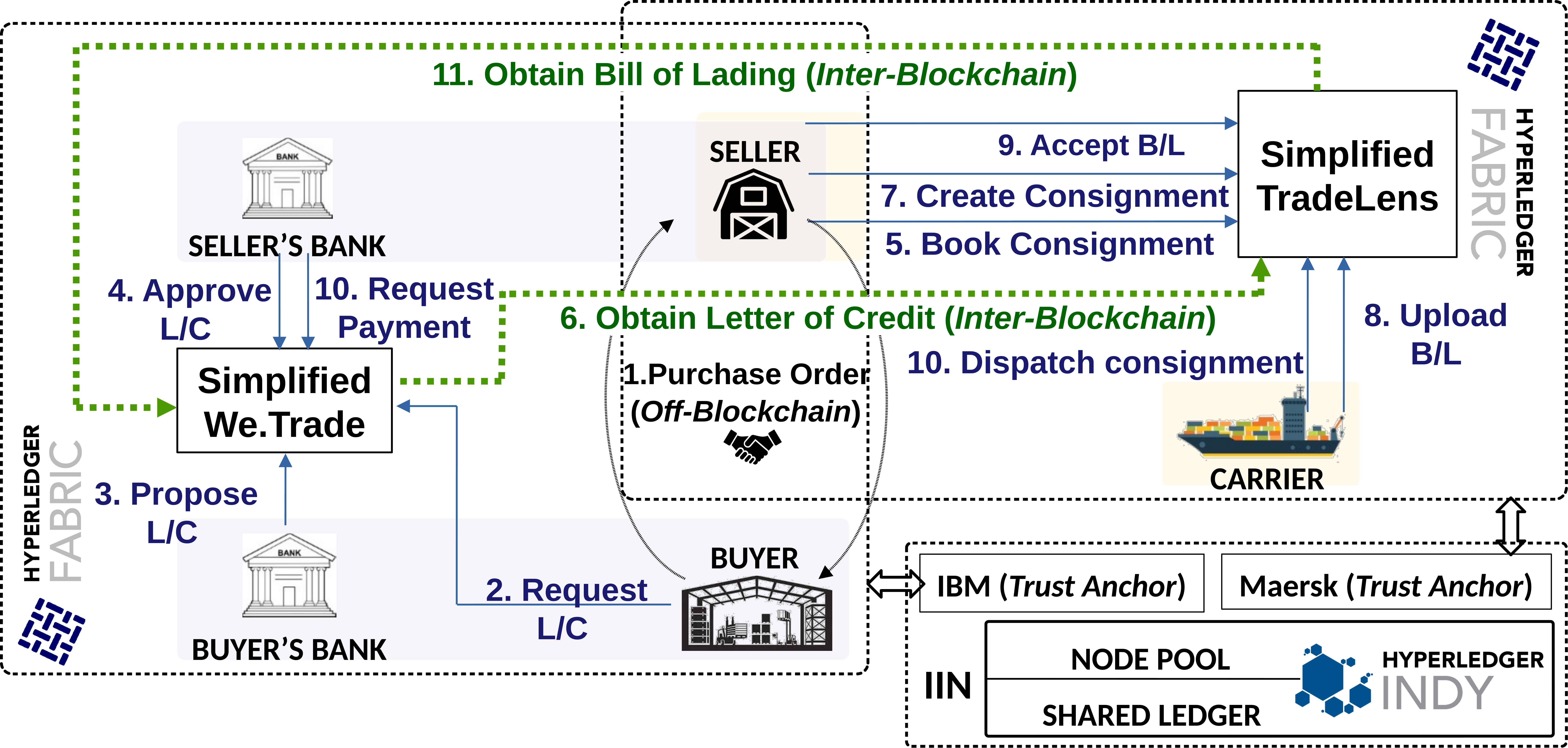}
 \caption{Simplified Cross-Network Trade Use Case}
 \label{fig:trade-flow-implementation}
\end{figure}

STL consists of a Seller and a Carrier organization, as illustrated in Figure~\ref{fig:trade-flow-implementation} each running a peer and a CA (serving as MSP). Consignments are created and dispatched in the workflow, with a bill of lading~\cite{billoflading} (B/L) recorded on ledger. SWT consists of a Seller and a Buyer organization, each with 2 peers and CAs, running a letter of credit~\cite{letterofcredit} (L/C) management workflow. SWT application clients include the same Seller that is a member of STL, a Buyer, and the Seller's and Buyer's banks. The interoperation steps are: (1) transfer of L/C from SWT to STL as a prerequisite for consignment creation and (2) transfer of B/L from STL to SWT for payment obligation enforcement. 

\subsection{Distributed Identity Infrastructure}
We used Hyperledger Indy to implement an IIN with trust anchors. Though other DID registries and DID providers exist, Indy is the most mature and offers all the features described in Section~\ref{sec:solution}. Indy maintains DID records on a public permissioned ledger shared by a pool of nodes running a consensus protocol~\cite{indy, aublin2013rbft}. Trust anchors called \textit{stewards} are bootstrapped into an Indy network within the genesis block, and these stewards can assign trust anchor privileges to other DID owners too: Indy supports TAs of the OIV and PMV categories out-of-the-box. DID records contain service endpoints, credential schemas, and authentication public keys (called \textit{credential definitions}). Real-world DIDs are called \textit{verinynms} (anonymous \textit{pseudonyms} are also supported), and a TA (OIV) can register a verinym on the Indy ledger using a special \textit{NYM transaction}.
TAs (including stewards) and IIN Agents (\textit{see below}) are implemented using the companion Hyperledger Aries~\cite{aries} framework, which enables confidential peer-to-peer communications among Agents and TAs.

For proof-of-concept, we deployed a single IIN: an Indy network bootstrapped with 4 independent Sovrin stewards~\cite{sovrin}. Two trust anchors were enrolled in the IIN by the stewards: one in the name of IBM to represent the SWT consortium and another in the name of Maersk to represent STL (see Figure~\ref{fig:trade-flow-implementation}). (Note that IBM and Maersk are initiators and major players in the real We.Trade and TradeLens networks respectively, and are therefore realistic sources of trust.) A single IIN with two trust anchors is sufficient to demonstrate operational mechanics in a proof-of-concept; in a production implementation, we will likely have more diversity but the mechanisms used will be identical to what we demonstrate here. The SWT Seller and Buyer register verinyms with, and obtain Membership VCs from, the IBM anchor in the IIN; likewise, the STL Seller and Carrier register and obtain theirs from the Maersk anchor.

\subsection{Fabric Network Organizations and Identity Providers}
In a Fabric network, identity is independently managed within an organization by one or more \textit{membership service providers} (MSPs)~\cite{fabric-msp}, implemented as a set of Fabric CA Servers~\cite{fabric-ca-server} (CA: \textit{Certificate Authority}). Multiple root and intermediate CAs can exist within an organization, creating trust chains. Each peer or transaction-submitting client \textit{enrolls} with an MSP (one of the Fabric CA servers) in their organization to obtain a unique identity and X.509 certificates for transaction signing. Organizations are then linked together on a channel when a configuration block containing their respective MSPs' root and intermediate CA certificates is appended to that channel's blockchain.

Every valid transaction in a block must carry a set of peer signatures that satisfies an \textit{endorsement policy}. Likewise, in a data-sharing instance, any data shared with an external network can only be deemed valid if it carries a set of peer signatures that satisfies a \textit{verification policy}. But proof verification (i.e., signature validation) requires the destination network to possess the source network's organization list and the certificates of its MSPs. Below we show how IIN Agents embedded within a Fabric network enable proof verification by fetching certificates and creating identity records on the ledger using the CMDAC contract.

\begin{figure}[t]
\centering
 \includegraphics[width=0.44\textwidth]{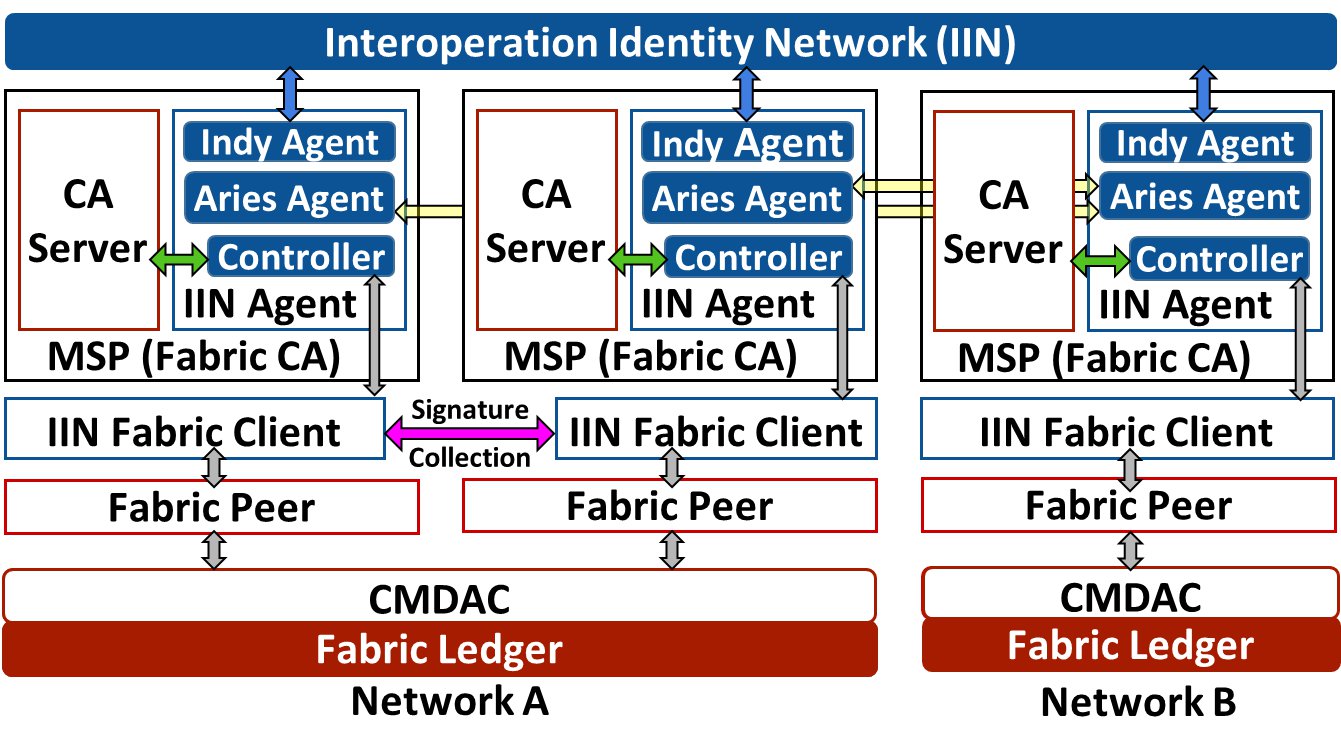}
 \caption{IIN Components and Connections for Fabric}
 \label{fig:implementation}
\end{figure}

\subsection{IIN Agents within a Fabric Network}

An IIN Agent represents an organization outside its network, and hence is designed to be an extension of that organization's MSP. 
An organization typically uses a single MSP in production, but if multiple MSPs are used, representing organizational units, we can use a different Agents for each.
Logically, the Agent functionality ought to be performed by a Fabric CA Server.
Rather than modifying the Fabric CA code, for implementation and deployment convenience, our IIN Agent is built as a decoupled service exposing an API for communication with IINs (Indy networks), IIN Agents of other local network organizations, and IIN Agents of foreign networks' organizations (see Figure~\ref{fig:implementation}). The IIN Agent also has client privileges and can submit transactions to the CMDAC. 

An IIN Agent is composed of three modules. An \textit{Indy Agent} built using with \texttt{indy-sdk}\cite{indy-sdk} is used to connect to an IIN's Indy pool and query DID/credential information. An \textit{Aries Agent}, implemented using \texttt{ACA-Py}\cite{aca-py}, is used to communicate with IIN trust anchors for verinym, VC, and VP requests. It uses DID service endpoints and credential definitions (authentication keys) for encrypted communication. While handling VCs and VPs, the \textit{Aries Agent} calls the \textit{Indy Agent} to fetch/update credential schema and definitions, and revocation lists. The \textit{Controller}, implemented using \texttt{Node.js} orchestrates the identity initialization, exchange and validation flow as described in Section~\ref{subsec:messageflow}. It is responsible for fetching its associated MSP's latest root and intermediate certificates for sharing with foreign networks. These three modules run within a single Docker container. Lastly, an \textit{IIN Fabric Client} application built on the \texttt{Fabric SDK}~\cite{fabric-sdk} updates foreign networks' identities and certificates on the channel ledger using CMDAC transactions. This app, running within its own container, takes transaction requests from the \textit{Controller} and executes an application level signature collection flow (see Section~\ref{subsec:sig-flow}) before invoking the CMDAC.

\subsection{Protocol: Syncing Foreign Identities through Consensus} \label{subsec:sig-flow}

Our protocol implementation follows the steps described in Section~\ref{subsec:messageflow} verbatim except for the final step in which a foreign network's identities are recorded onto the local ledger. This step requires DLT-specific mechanisms, and we will show how they were implemented in Fabric. As an example, we consider the scenario where SWT is trying to fetch and sync the certificates of the Carrier organization in STL.

Recording the STL Carrier's certificates on the ledger creates a shared truth for the entire SWT network, enabling its peers to refer to those certificates in a data transfer session, either for access control or proof verification. Allowing the CMDAC to directly query IINs or foreign networks' agents to fetch these certificates would create a non-determinism hazard. Therefore, we use an application-level flow involving IIN Agents for deterministic invocation of CMDAC.

The SWT Buyer IIN Agent initiates this flow by validating the STL Carrier's Membership VC with the IIN, and subsequently fetches the Carrier's certificates from the  Carrier's IIN Agent (whose service endpoint is part of the Carrier's DID record). The Buyer IIN Agent then sends these certificates to its Fabric Client app, which then prepares a \textit{signature collection request}, and sends it to the Fabric Client of the SWT Seller IIN Fabric Client app. The latter validates the Carrier's certificates after consultation with their own IIN Agents (who have presumably fetched those certificates too), and then counter-signs the request on behalf of its organization. The Buyer aggregates the responses (only one here) and submits it as input in a CMDAC transaction. The chaincode checks for the presence of valid signatures from every SWT organization (here: Buyer and Seller) before approving the update of the STL Carrier's identity state on the ledger. Note that this update is idempotent, and can be carried out concurrently by the IIN Agents of both Buyer and Seller. Also, if the Buyer's and Seller's IIN Agents' copies of Carrier certificates are not in sync, the signature collection flow will fail and must be retried.

This protocol replaces the naive implementation in Abebe et al~\cite{abebe2019enabling} where organizational identities and root and intermediate certificates were fetched out-of-band manually.
The identity plane exchange we have demonstrated makes the process more secure and consensus-based. Further, any changes in organizational memberships or certificate belonging to an organizations can be determined and synced automatically through periodic queries made to the IIN registry (for membership and revocation lists) or whenever a proof validation fails because of expired certificates.

	\section{Analysis}\label{sec:analysis}

We now analyse our system's ease of use and extensibility, and discuss its limitations with a view to future improvements. 

\subsection{Generality and Flexibility}\label{sec:flexibility}
Our design consists of two distinct sets of components: \emph{1) Distributed Identity Infrastructure} shared by interoperating networks but existing outside them, namely the IINs, and \emph{2) Network Identity Managers} components that lie within networks, namely the IIN Agents. IINs are built using state-of-the-art industry standards (Indy, DID, VC), though the specification is independent of a specific technology.
IIN Agents are DLT-specific and decentralized within a network, lying within the scope of a network's participant/organization. In fact, different organizations may implement their own versions of IIN Agents and replace them independently; an Agent just needs to expose the API we have specified earlier in this paper.

\subsection{Security}
We evaluate the security of our protocol against the standard CIA triad model~\cite{whitman2011}. \emph{Confidentiality:} The DIDs of network participants are themselves public by necessity, as the DID Registry (IIN) is a public permissioned network. But a DID by itself only reveals the existence of an organization that participates in a network without revealing anything else about that organization, like membership information,
which are known only to identity owners and IIN trust anchors. Also, the intra-network certificates are shared point-to-point among IIN Agents on a need basis and are thus kept confidential from everyone outside the interoperating networks.
\emph{Integrity:} Identities are registered in an IIN using a fault-tolerant consensus protocol (Indy typically uses RBFT\cite{aublin2013rbft}), thereby ensuring a high level of integrity. Trust anchors maintaining Memberlist VCs are assumed to have reputations and are trusted by organizations belonging to a consortium; further, identities and VCs are attested by signatures using keys registered in IINs. Integrity violations are therefore unlikely but can be easily detected, allowing organizations to select more trustworthy anchors.
\emph{Availability:} The availability of identity records depends on the size of the IIN; the more the number of nodes in an Indy pool, the higher the availability. An IIN Agent or an IIN Trust Anchor by itself can be a point of failure, but this can be mitigated by adding redundancy.

\subsection{Ease of Extensibility}

\subsubsection{Network Identity Managers}
An IIN Agent runs as part of a network, but only portions of it needs to have a DLT-specific implemenation. Examining the protocol steps in Section~\ref{subsec:messageflow}, we see that step A involves the Agent communicating with IINs and Trust Anchors using a standard API. Similarly, steps B and C involves communication between IIN Agents across network boundaries, again using a standard interface.
Only Step D, which involves updating the local ledger via a smart contract transaction must be DLT-specific. Hence, an IIN Agent can be mostly built using off-the-shelf components. The transaction submission component must be DLT-specific, as was the IIN Fabric Client described in Section~\ref{sec:implementation}. The equivalent of this in Corda would be a CorDapp~\cite{cordapp} and in Hyperledger Besu would be a Dapp~\cite{dapp}. 

\subsubsection{Distributed Identity Infrastructure}
Though our IINs are implemented using Indy and IIN trust anchors using Aries, their specifications and interfaces are based on W3C standards for DIDs and VCs and VPs. Therefore, they can easily be ported to other verifiable data registries that follow the same standard (e.g. Sidetree \cite{sidetree}).

\subsection{Limitations}
A trust anchor representing a consortium or unilaterally issuing real-world DIDs to organizations is the only centralized component in our implementation. But further decentralization is possible by requiring more than one TA to vouch for a network participant; e.g., using a smart contract in the IIN. Collaborative models, where TAs (e.g. representing Fabric MSPs) corroborate each other using signatures can also enhance safety and liveness of identity plane protocols.

Decoupling of network participants identities from their IIN identities presents another challenge: syncing the two sets to ensure that networks possess up-to-date info for data plane operations. In general, this only affects liveness and not safety, because proof verification failures can be handled by re-synchronizing identities using strategies like polling to event triggers. While polling may be slightly inefficient from a communication standpoint, it provides a higher level of assurance (depending on the polling interval). Additional watchdogs may be needed to handle all cases, in case of events, in both the identity and data planes.

	\section{Related Work}\label{sec:relatedork}

Networks built on Corda~\cite{corda} can transact states representing data and assets with each other via the Corda Network~\cite{cordanetwork}, a global publicly-available
network that uses a common root of trust for identity. Though a
consortium of nodes may optionally choose to deploy a segregated
network with its own trust root for privacy and
confidentiality, it will be unable to communicate directly with
the rest of the global network unless it merges with it.
Unlike our DLT-agnostic architecture, which enables independent segregated networks to interoperate in a privacy-preserving manner without relying on a common network, the Corda Network suffers from a key limitation: it is restricted to the Corda protocol and
doesn't allow integration with other DLT protocols like 
Fabric~\cite{fabric} and Besu~\cite{besu}.

In the permissionless domain there are a range of efforts attempting
to address identity which include naming services such as
Ethereum Name Service (ENS)~\cite{ens} and Polkadot's naming system~\cite{polkadotidentity}, as well as a number of solutions
based on the Decentralized Identity Framework such as
uPort~\cite{uPort} and Ontology~\cite{ontio}. However, these systems
are either designed to simplify user experience in public networks,
such as addressing an entity with a user-friendly name instead of an
arbitrary byte string, or to provide a user-facing identity solution
for creating and sharing credentials. These systems don't address the
general problem of resolving and verifying identity issued by
different ledgers for enabling cross-ledger communication. 

Hyperledger Cactus~\cite{cactus} leaves networks autonomous and in control of their interactions with other networks, but currently relies on manually sharing network identity information. 

To the best of our knowledge, ours is the first attempt at formalizing the separation of proof and identity concerns, and presenting an identity exchange solution (based on SSI) that adheres to blockchain tenets of decentralized trust.


	\section{Conclusion and Future Work}\label{sec:conclusion}

Interoperation for data sharing between permissioned blockchain networks running related business processes requires the networks to have the ability to identify each others' participants and validate their claims/proofs. We have described a way of reasoning about such protocols, separating identity concerns from data and policy concerns into a different communication plane. To give networks the ability to prove memberships, cross-validate identities, and share certificates, we have designed a DLT-agnostic architecture and protocols based on self-sovereign identity and verifiable credential concepts. A proof-of-concept implementation was demonstrated, linking two Hyperledger Fabric networks. 
This consisted of an identity registry (IIN) built on Hyperledger Indy and agents built on Hyperledger Aries exchanging certificates across network boundaries in a peer-to-peer manner. In the future, we intend to demonstrate compatibility with Corda and Hyperledger Besu, distribute the trust currently invested in IIN trust anchors, and conduct performance evaluations.

\section*{Acknowledgement}
We thank Petr Novotny (IBM T.J. Watson Research Center) for playing a key role in crafting the idea that led to this work.
	
	\bibliographystyle{IEEEtran}
	
	\bibliography{biblio1}

\end{document}